\documentclass[twocolumn,10pt,longbibliography,nofootinbib]{revtex4-1}

\usepackage{amsfonts,amssymb}
\usepackage[sumlimits,intlimits]{amsmath}
\usepackage{graphics}
\usepackage{graphicx}
\usepackage{mathrsfs}
\usepackage{textcomp}
\usepackage{verbatim}
\usepackage{hyperref}    
\usepackage{bm}          

\newcommand{\be}{\begin{equation}}
\newcommand{\ee}{\end{equation}}
\newcommand{\nhet}{N_\text{het}}
\newcommand{\Tt}{\widetilde{T}}

\begin{document}

\title{Mathematical toy model inspired by the problem of the adaptive origins of the sexual orientation continuum}

\author{Brian Skinner}
\affiliation{Massachusetts Institute of Technology, Cambridge, MA  02139 USA}

\date{\today}

\begin{abstract}

Same-sex sexual behavior is ubiquitous in the animal kingdom, but its adaptive origins remain a prominent puzzle.  Here I suggest the possibility that same-sex sexual behavior arises as a consequence of the competition between an evolutionary drive for a wide diversity in traits, which improves the adaptability of a population, and a drive for sexual dichotomization of traits, which promotes opposite-sex attraction and increases the rate of reproduction.  This tradeoff is explored via a simple mathematical ``toy model."  The model exhibits a number of interesting features, and suggests a simple mathematical form for describing the sexual orientation continuum.  

\end{abstract}

\maketitle

\section{Introduction}

When a particular behavior or trait is widespread across a group of animals, its origin is usually explained in terms of the fitness advantage that it confers.  Such explanations attempt first to understand how the fitness of the animal population has a dependence on the degree to which it exhibits a given trait.  It is then assumed that the processes of evolution and natural selection bring the population close to the point of maximal fitness.  (See, for example, Refs.\ \onlinecite{smith_optimization_1978, parker_optimality_1990} for a review.)

Given this paradigm, the prevalence of same-sex sexual behavior in the animal kingdom has presented something of a puzzle.  Same-sex sexual behavior is ubiquitous across the animal kingdom, and has been cataloged in hundreds of animal species in ways that range from same-sex courtship and copulation to long-term pair bonding and parenting.  (See, for example, Ref.\ \onlinecite{bagemihl_biological_1999} for an extensive review.)  This ubiquity suggests the possibility that same-sex behavior is associated with some kind of fitness advantage.  The nature of this advantage, however, remains poorly understood, and is a source of considerable scientific debate.  The puzzle is particularly pronounced because same-sex attraction ostensibly has a significant cost, in the sense that it can reduce the frequency of mating between opposite-sex pairs, and thereby lower the rate of reproduction.

A number of hypotheses have been proposed to explain the origin of same-sex sexual behavior in animals.  These are reviewed, for example, in Refs.\ \onlinecite{poiani_animal_2010, bailey_same-sex_2009}, but a few of the more prominent hypotheses are briefly listed as follows.  One hypothesis is that such behaviors arise primarily because of their role in maintaining social bonds, alliances, and dominance hierarchies among members of the same sex.  Another possible mechanism is that same-sex courting or mating provides ``practice" that improves the odds of success in later mating attempts with the opposite sex.  Some studies have also considered the ``kin selection" hypothesis, which posits that same-sex sexual behavior in one individual provides a genetic advantage to the individual's siblings, and on the whole provides an advantage to the family genetic line.  Finally, there are genetically-motivated hypotheses, such as the idea that genes promoting same-sex sexual behavior in a homozygous state may confer a fitness advantage when in a heterozygous state, or the idea that an allele promoting same-sex sexual behavior in one sex may increase the fitness of the opposite sex.  (Table 2 of Ref.\ \cite{bailey_same-sex_2009} provides a summary of these and other hypotheses, along with further references.)

The purpose of this paper is to define and consider an interesting mathematical problem that can be said to describe a different potential mechanism for the adaptive origins of same-sex sexual behavior.  Central to this proposed mechanism are two ideas: first, that having a diversity of traits among a given group confers a fitness advantage, and second, that the sexual attraction of one individual to another is determined by the traits of the other, rather than by their genetic sex.  These two ideas together imply that the breadth of traits present within a given sex is pulled in opposite directions by two competing factors.  On the one hand, the unpredictable environment favors a wide distribution of traits.  On the other hand, the sexual nature of reproduction favors a dichotomizing of traits according to each individual's biological sex.  Such a dichotomy promotes opposite-sex attraction, thereby increasing the number of offspring.  The purpose of this paper is to explore the idea that a balance exists between these two factors that naturally leads to a finite degree of same-sex sexual attraction.  

It should be stated up front that this paper is not intended to be taken as a realistic model for explaining the sexual behaviors or sexual orientations of any particular animal group.\footnote{Of course, this paper also does not attempt to analyze the relative advantages of sexual versus asexual reproduction; only sexual reproduction is considered in the model presented here. Questions about the origin of sex have been studied elsewhere (for example, in Refs.\ \onlinecite{bernstein_origin_1984, kondrashov_asexual_1994, agrawal_sexual_2001, siller_sexual_2001}), and remain a prominent research topic.}  Instead, I focus only on a simple mathematical problem, which represents a minimal description of a possible tradeoff between diversity of traits and sexual dichotomization.  Further, this analysis considers only the distribution of traits and preferences that maximize the expected fitness of the population as a whole.  Whether this kind of optimum can be expected to be produced by evolution and natural selection is a delicate question, and depends on the mechanisms by which sexual traits and sexual preferences are (or are not) inherited \cite{smith_optimization_1978, parker_optimality_1990}.  These questions are not considered here, and as such this paper is best read as merely an interesting mathematical problem that is \emph{inspired} by the question of the origins of diversity in animal sexual behavior.  The hope is that this problem, and its solution, can inspire future discussion and more accurate models.

The remainder of this paper is devoted to proposing and exploring a simple ``toy model", which considers the optimal distribution of a single trait among the population.  The distribution is completely determined by a single parameter $t$ that describes the relative importance of phenotypic variation for the species fitness. The value of the parameter $t$ determines both the distribution of traits among the population and the prevalence of same-sex pairing, both of which can be described analytically.  The model exhibits a number of interesting mathematical features, including a series of bifurcations in the trait distribution and in the distribution of sexual orientations as a function of $t$.  At small $t$, both distributions acquire a simple mathematical form.  Results from the model are discussed in the context of data on human sexual orientation.

\section{Model}
\label{sec:model}

\begin{figure}[htb]
\centering
\includegraphics[width=0.45 \textwidth]{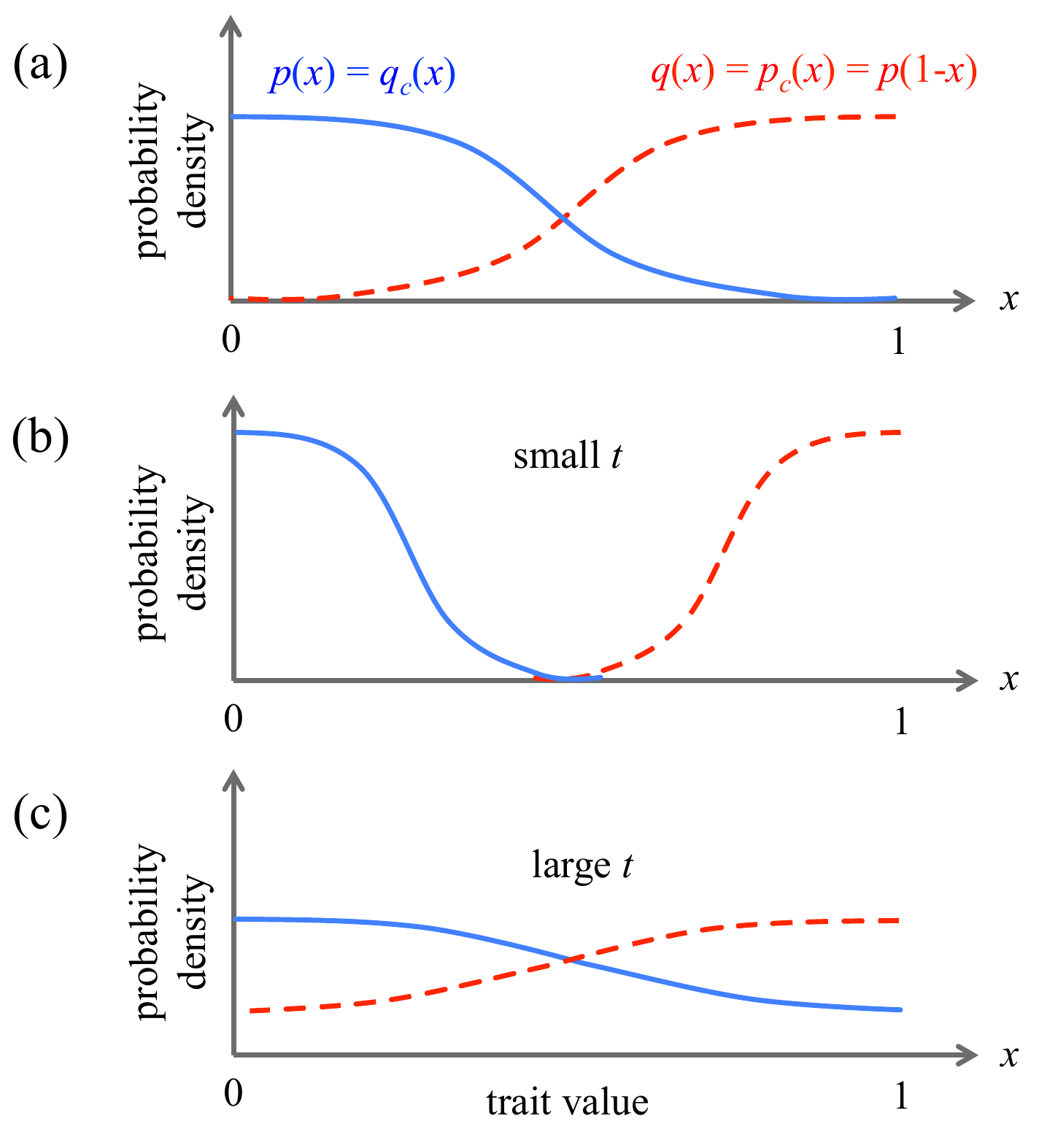}
\caption{(a) Schematic depiction of the four relevant distributions of trait value and trait preference: $p(x)$, the distribution of traits possessed by males; $q(x)$, the distribution of traits possessed by females; $p_c(x)$, the distribution of traits desired by males in a mate; and $q_c(x)$, the distribution of traits desired by females in a mate.  Within the model, all four can be related to a single distribution $p(x)$, which is to be optimized. (b) When the parameter $t$ is small, the optimal distributions are such that $p(x)$ and $q(x)$ have very little overlap, and the number of offpsring is maximized.  (c) When $t$ is large, a broader distribution of traits is favored, and consequently there is significant overlap between the male and female trait distributions, resulting in a relatively high rate of same-sex pairing. }
\label{fig:ps}
\end{figure}

In the toy model that is the subject of this paper, it is imagined that all individuals are characterized by a single trait whose value $x$ ranges from $0$ to $1$.  Suppose, for concreteness, that females tend to have values of $x$ that are closer to $1$, while males tend to have values of $x$ that are closer to $0$.  Under this description, each sex is characterized by two probability density functions: one describing the probability of possessing a certain trait value $x$, and the other describing the probability of desiring a trait value $x_c$ in a mate.  The distributions of the trait value $x$ are denoted $p(x)$ and $q(x)$ for males and females, respectively.  The distribution of the desired trait value $x_c$ is denoted $p_c(x_c)$ for males and $q_c(x_c)$ for females.  The four distributions are summarized graphically in Fig.\ \ref{fig:ps}.   It is assumed that $x$ and $x_c$ are independent variables, so that the trait value $x_c$ that an individual desires in a mate is independent of the trait value $x$ possessed by the individual itself.

In principle, these four distributions can be completely distinct from each other.  However, in order to simplify the model I introduce the following two assumptions.  The first assumption is that there is a symmetry between the two sexes, such that each sex is equivalent to the other under a redefinition of the value of the trait $x \rightarrow 1-x$.  In other words, in terms of their traits and preferences, the two sexes are taken to be ``mirror images" of each other, so that $q(x) = p(1-x)$ and $q_c(x_c) = p_c(1-x_c)$.  The second assumption is that the number of individuals possessing trait value $x$ is equal to the number of individuals desiring the trait value $x$ in a partner.  This assumption guarantees that there is ``someone for everyone", and is equivalent to the conditions that $p_c(x) = q(x)$ and $q_c(x) = p(x)$.  These two assumptions together imply that there is only one relevant distribution $p(x)$ for describing the two sexes, and that all others can be related to it by $p_c(x) = q(x) = p(1-x)$ and $q_c(x) = p(x)$ (see Fig.\ \ref{fig:ps}). 

Now consider a population consisting of a very large number $N$ of individuals, and suppose that the individuals all become paired with each other in such a way that every individual's desire for the trait value of their partner is satisfied.  The proportion of heterosexual pairings that result from this process can be calculated as follows.      

Consider two different trait values $x_1$ and $x_2$.  One can now define two groups of individuals: (1) those who possess trait value in the infinitesimal interval $(x_1, x_1 + dx_1)$ and desire trait value $(x_2, x_2 + dx_2)$ in a partner, and (2) those who similarly possess $x_2$ and desire $x_1$.  These two groups are referred to as ``group 1" and ``group 2", respectively.  The number of males in group 1 is given by $M_1 = N \cdot p(x_1) dx_1 \cdot p_c(x_2) dx_2$.  Similarly, the number of females in group 1 is $F_1 = N \cdot q(x_1) dx_1 \cdot q_c(x_2) dx_2$.  For group 2, one can likewise define the number of males and females as $M_2 = N \cdot p(x_2) dx_2 \cdot p_c(x_1) dx_1$, and $F_2 = N \cdot q(x_2) dx_2 \cdot q_c(x_1) dx_1$, respectively.  Because of the symmetry of the distributions $p(x)$ and $q(x)$, the total number of individuals $M + F$ is the same in both groups.  One can therefore pair the two groups in such a way that each individual in group 1 is paired with an individual in group 2.  If these pairings are selected at random, then the proportion of heterosexual pairings is $(M_1 F_2 + F_1 M_2)/(M+F)^2$, and the number of heterosexual pairings between the two groups is $d\nhet(x_1, x_2) = (M_1 F_2 + F_1 M_2)/(M+F)$.  To find the total number of heterosexual pairings across the entire population, one can integrate $d\nhet(x_1, x_2)$ over all values of $x_1, x_2$.  Inserting the expressions for $M_{1,2}$ and $F_{1,2}$ gives
\be 
\nhet = \frac{N}{2} \int_0^1 \int_0^1 \frac{ \left[ (p(x_1) p(1 - x_2) \right]^2 + \left[ (p(1-x_1) p(x_2) \right]^2 }{p(x_1) p(1-x_2) + p(1-x_1) p(x_2) } dx_1 dx_2 .
\label{eq:Nhet}
\ee 
The value of $\nhet$ is maximized when the distributions of possessed traits and desired traits, $p(x)$ and $p(1-x)$, have zero overlap [i.e., when $p(x) p(1-x) = 0$ everywhere].  In this case all pairings are heterosexual, $\nhet = N$.  If each heterosexual pairing produces $b$ offspring on average, then the number of individuals in the next generation is $b\nhet $.  

On the other hand, one may expect finite overlap between $p(x)$ and $p(1-x)$ in situations where there is a fitness advantage conferred by each sex having a wide diversity in traits.  In particular, one can define the trait \emph{entropy} of the next generation as
\be 
S =  -b \nhet \int_0^1 p(x) \ln p(x) dx.
\label{eq:S}
\ee 
Equation (\ref{eq:S}) is equivalent to the Shannon entropy $s$ of the distribution $p(x)$, multiplied by the number of individuals in the population.  The entropy $S$ is maximized when $p(x) \equiv 1$, \textit{i.e.}, when every trait value is equally likely for each individual, regardless of sex.  Presumably, when the environment is such that there is pressure to produce offspring and also pressure to maintain a diversity of traits, the distribution $p(x)$ will reach a steady-state that involves a tradeoff between maximizing the number of offspring and maximizing the entropy of the trait distribution [see Fig.\ \ref{fig:ps}(b) and (c)].

To model that tradeoff, I introduce a generic fitness function $F$ that consists of a term proportional to the total offspring number plus a term proportional to the trait entropy.  In other words, the proposed fitness function is
\be 
F = u_0 \nhet  + T_0 S,
\nonumber
\ee 
where $u_0$ and $T_0$ are constants that arise from environmental pressures and are independent of the distribution $p(x)$.  Dividing both sides of this equation by $u_0 N b$ one arrives at a renormalized fitness function $f = F/(u_0 N b)$ that is a function of only a single parameter $t$:
\be 
f = n \left( 1 + t s\right).
\label{eq:f}
\ee 
Here, $n = \nhet/N$ [see Eq.\ (\ref{eq:Nhet})] and $s = S/(b \nhet)$ [see Eq.\ (\ref{eq:S})] are functionals of the trait distribution $p(x)$, and $t = T_0/u_0$ is a dimensionless ``entropy parameter" that characterizes the relative importance of trait diversity.  (In this sense, $t$ plays a role similar to that of the temperature in the Helmholtz free energy of statistical physics.)  When $t = 0$, the optimum distributions have no overlap between male and female traits, and all pairings are heterosexual ($n = 1$).  When $t \rightarrow \infty$, on the other hand, the population fitness is optimized by $p(x) \equiv 1$, and heterosexual and homosexual pairings are equally likely ($n = 1/2$).  

In the remainder of this paper, results are presented for the distribution $p(x)$ at different values of the parameter $t$.  The primary tool used for finding the optimal $p(x)$ is a numerical Monte Carlo algorithm, which is described in the Appendix.  Briefly, this algorithm divides the trait interval $[0, 1]$ into discrete points $x_i$, and makes an initial guess for the function $p(x_i)$.  The values of $p(x_i)$ are then optimized by making random deviations from the initial guess, and then evaluating the corresponding change to the population fitness $f$.  Changes are kept or discarded according to the Metropolis algorithm, and the procedure is iterated until a convergent solution is found.

Once the distribution $p(x)$ is known, one can also examine the corresponding distributions of ``sexual orientation" $\theta$, which is defined as the probability of a given individual pairing with a same-sex rather than an opposite-sex partner.  In particular, for an individual (say, a male) that prefers a trait value $x_c$ in a partner, one can define the orientation $\vartheta(x_c)$ of the individual as the proportion 
\be 
\vartheta(x_c) = \frac{p(x_c)}{p(x_c) + q(x_c)} = \frac{p(x_c)}{p(x_c) + p(1 - x_c)}
\ee
of same-sex individuals among the group to which the individual is attracted.  One can also define a probability density function for $\theta$ as 
\be 
P(\theta) = \int p_c(x_c) \delta\left(\theta - \vartheta(x_c) \right) d x_c,
\label{eq:orientationdist}
\ee 
where $\delta(x)$ is the Dirac delta function.
In the following section, results are presented for both the trait distribution $p(x)$ and the orientation distribution $P(\theta)$ as a function of the entropy parameter $t$.

\section{Results}
\label{sec:results}

\begin{figure}[htb]
\centering
\includegraphics[width=0.48 \textwidth]{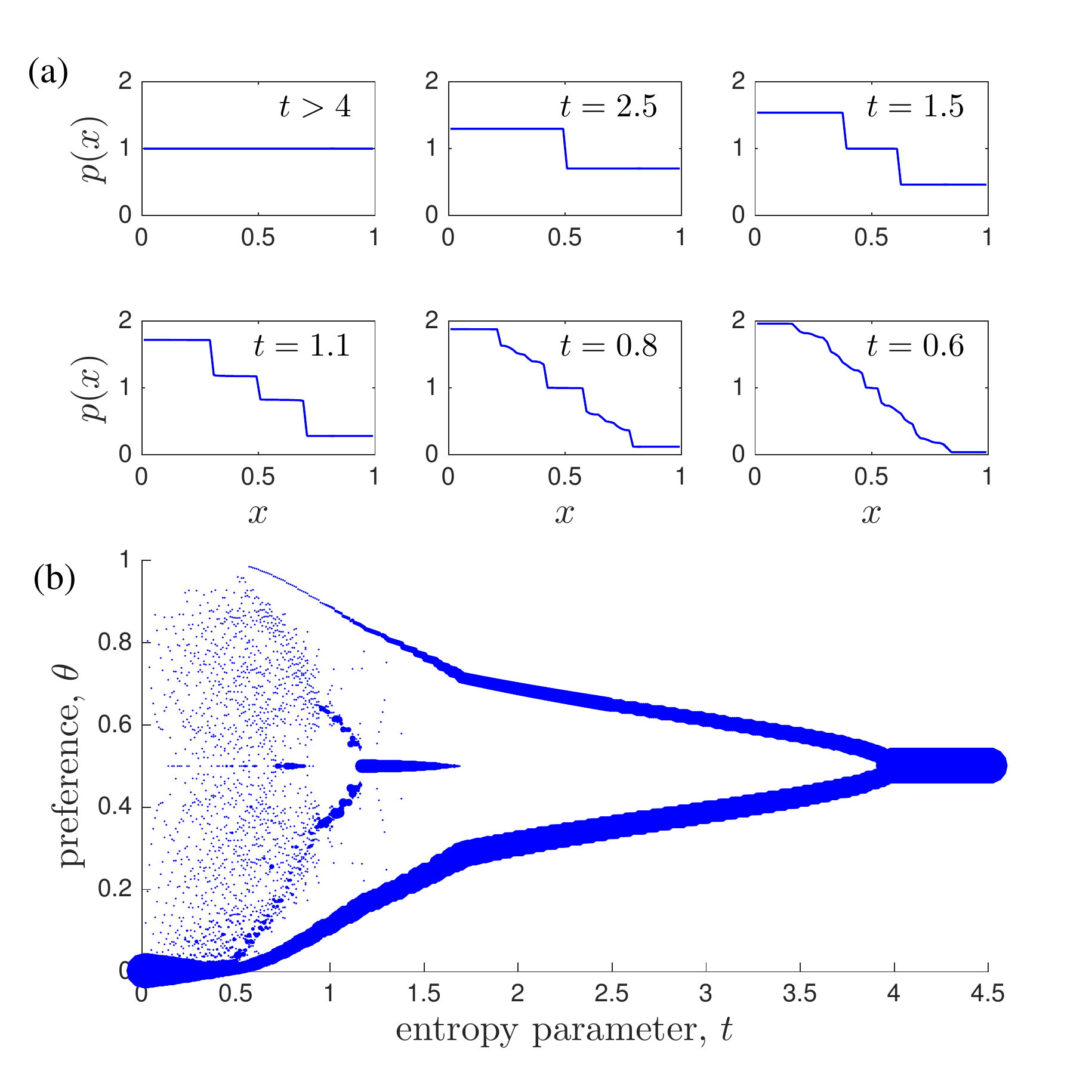}
\caption{(a) Evolution of the trait distribution $p(x)$ with decreasing entropy parameter $t$.  Different subplots are labeled by the corresponding value of $t$.  (b) Plot of the values of sexual orientation $\theta$ observed in the population for different values of $t$.  Points represent values of $\theta$ arising from the trait distribution $p(x)$.  The size of the points indicates the relative abundance of that orientation.}
\label{fig:pxplot}
\end{figure}

When the entropy parameter is large, $t \gg 1$, the trait distribution becomes flat, $p(x) \equiv 1$, which maximizes the trait entropy at the cost of reducing the total number of offspring by $50\%$.  In fact, the optimal distribution is precisely equal to $p(x) \equiv 1$ for all values of $t \geq 4$.  Only at $t < 4$ do traits begin to specialize according to sex.  At $t$ slightly smaller than $4$, the distribution $p(x)$ acquires a step-like shape, with traits corresponding to $x < 1/2$ being more prevalent in males, and traits with $x > 1/2$ being more prevalent in females.  This transition is depicted in Fig.\ \ref{fig:pxplot}(a).

One can describe the transition at $t = 4$ analytically by writing the distribution $p(x)$ as 
\[ p(x) = \begin{cases} 
      1 + c, & x < 1/2 \\
      1 - c, & x > 1/2 
   \end{cases} ,
\]
where $c$ is a parameter to be determined.  Inserting this distribution into Eqs.\ (\ref{eq:Nhet}) and (\ref{eq:S}), one can evaluate the frequency $n$ of opposite-sex pairing as $n = 3/2 - 1/(1+c^2)$, and the trait entropy as $s = [(1+c) \ln(1+c) + (1-c) \ln (1-c)]/2$.  Expanding these expressions to lowest order in $c$ gives a fitness function $f = 1/2 + c^2(1 - t/4) - c^4 t/2$, which is minimized when
\be 
c = \sqrt{ \frac{4-t}{4t} }.
\ee 
In other words, at $t \geq 4$ the optimal fitness is provided when $c = 0$, and the trait distribution is uniform.  At $t < 4$, on the other hand, there emerges a difference in trait distributions between the two sexes, with a magnitude $c$ that grows as $\sqrt{4 - t}$.

This splitting also has an implication for the distribution of sexual orientations, $P(\theta)$.  At $t > 4$, when the trait distribution is uniform, all individuals have orientation $\theta = 1/2$, since there is no sexualization of traits.  When $t$ is lowered below $4$, on the other hand, there emerge two classes of orientation: $\theta = (1 \pm c)/2$.  The former class (with a majority preference for same-sex partners) comprises a smaller proportion $(1 - c)/2$ of the population.  The latter class (with a majority preference for opposite-sex partners) comprises a larger proportion $(1 + c)/2$.  In other words, the distribution of orientation $P(\theta)$ is such that $P(\theta) = \delta(\theta - 1/2)$ at $t > 4$, while at $t$ slightly less than $4$ one has $P(\theta) = \frac{1+c}{2} \delta\left(\theta - \frac{1-c}{2}\right) + \frac{1-c}{2}\delta \left(\theta - \frac{1+c}{2}\right)$.  This bifurcation of the orientation distribution is depicted in Fig.\ \ref{fig:pxplot}(b).  

As $t$ is reduced even further, the trait distribution undergoes a sequence of additional splittings, as illustrated in Fig.\ \ref{fig:pxplot}(a).  At $t \lesssim 1.7$, for example, the two-step structure of the trait distribution undergoes a transition to a three-step structure.  In terms of the orientation distribution, one can say that a third class of individuals with orientation $\theta = 1/2$ emerges in between the other two, and $P(\theta)$ is a sum of three Dirac delta functions.  At $t \lesssim 1.17$, this three-class structure transitions to a four-class structure, and as $t$ is reduced an increasingly large number of classes emerge.  

\begin{figure}[htb]
\centering
\includegraphics[width=0.45 \textwidth]{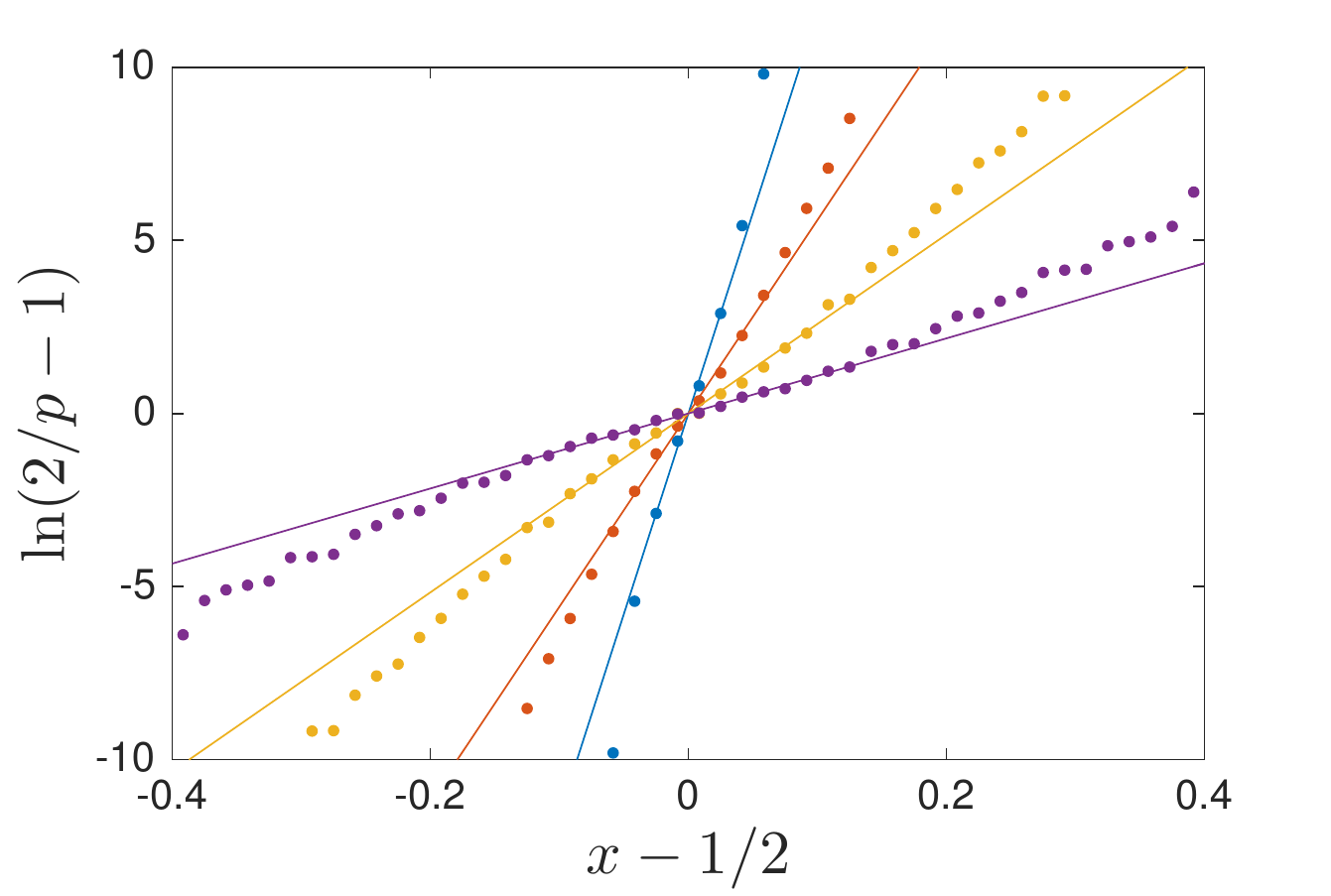}
\caption{Plot of the trait distribution $p(x)$ in the form $\ln(2/p - 1)$ versus $x-1/2$.  Plotted in this way, the ``Fermi function" form, Eq.\ (\ref{eq:Fermi}), corresponds to a straight line with zero intercept and a slope equal to $1/\Tt$.  In order of decreasing slope, the different curves correspond to $t = 0.05$, $t = 0.1$, $t = 0.2$, and $t = 0.4$.  The points show numerical results and the lines are the analytical solutions of Eqs.\ (\ref{eq:Fermi}) and (\ref{eq:T}), with no fitting parameters.}
\label{fig:fermi}
\end{figure}

When $t$ becomes small, $t \ll 1$, the distribution $p(x)$ has so many steps that it closely approximates a continuous function.  As shown in Fig.\ \ref{fig:fermi}, in this limit this function closely matches the form
\be 
p(x) \simeq \frac{2}{ 1 + \exp\left[(x - 1/2)/\Tt \right] },
\label{eq:Fermi}
\ee
which is reminiscent of the Fermi function from quantum statistical mechanics.  The parameter $\Tt$, which for the Fermi function is related to the system temperature, is linearly proportional to the entropy parameter $t$ at small $t$.

To derive the relation between $\Tt$ and $t$, one can insert Eq.\ (\ref{eq:Fermi}) into Eqs.\ (\ref{eq:Nhet}) and (\ref{eq:S}).  Evaluating the corresponding integrals at small $\Tt$ gives $n \simeq 1 - \pi^2 \Tt^2$ and $s \simeq - \ln 2 + \pi^2 \Tt/3$.  The fitness function $f = n(1 + ts)$ is then minimized when
\be 
\Tt \simeq \frac{t}{6(1 - t \ln 2)}.
\label{eq:T}
\ee
This solution minimizes Eq.\ (\ref{eq:f}) to within a term of order $t^3$, suggesting that Eq.\ (\ref{eq:Fermi}) is exact in the limit $t \rightarrow 0$.

Equation (\ref{eq:Fermi}) also implies a specific, continuous form for the distribution of sexual orientations, $P(\theta)$.  In particular, evaluating Eq.\ (\ref{eq:orientationdist}) gives
\be 
P(\theta) = \frac{2 \Tt}{\theta}.
\label{eq:fermiorientation}
\ee 
Notice that, for any nonzero value of the entropy parameter $t$, the distributions of male and female traits always have finite overlap, and consequently there are no individuals with strictly heterosexual or homosexual orientation, $\theta = 0$ or $\theta = 1$.  Consequently, the distribution $P(\theta)$ should be considered to be defined only over the interval $[\theta_\text{min}, \theta_\text{max}]$, where $\theta_\text{min} = \vartheta(x_c = 1) = [1 + \exp(1/2\Tt)]^{-1}$ and $\theta_\text{max} = \vartheta(x_c = 0) = [1 + \exp(-1/2\Tt)]^{-1}$.  In this sense the probability distribution $P(\theta)$ is properly normalized, since $\int_{\theta_\text{min}}^{\theta_\text{max}} P(\theta) d\theta = 1$.

\section{Discussion}

In this paper I have considered a simple mathematical toy model for the tradeoff between sexual dichotomy of traits and trait diversity.  Among the more interesting features of the model are the series of sharp transitions in the trait distribution as the parameter $t$ is varied, and the ``Fermi function" shape of the distribution at small values of $t$.  Of course, the model has employed a number of fairly artificial assumptions, most notably the assumption of a single relevant trait that is defined on the interval $[0, 1]$.  Since this assumption is unlikely to be applicable to a real biological population, it may be difficult to find direct empirical comparisons to the trait distribution $p(x)$.

On the other hand, the model also makes specific predictions about the distribution of sexual orientation, which can in principle be observed.  For example, the model suggests that when the relative importance of trait diversity is high (or, equivalently, when the relative importance of producing a large number of offspring is low), the population can be divided into a small number of well-defined groups with similar sexual orientation.  As the environment is changed in such a way that trait diversity becomes less important, these groups split into a larger number of groups through a sequence of sharp transitions.  Finally, when the value of trait diversity is low, the distribution of sexual orientation becomes continuous and takes the form $P(\theta) \propto 1/\theta$.

In principle, some of these results can be tested empirically by measuring the frequency of same-sex versus opposite-sex mating or pairing for a large number of individuals across an animal population.  (Of course, one should be cautious about conflating the observed frequency of same-sex behaviors with the internal preference of an individual for same-sex partners.)  Unfortunately, I am unaware of any studies that present sufficient data to construct an empirical version of the distribution $P(\theta)$. 

To date, the vast majority of quantitative research about same-sex sexual behavior focuses on humans.  Some studies, beginning with the Kinsey reports,\cite{kinsey_sexual_1998, kinsey_sexual_1998-1} have made an effort to assess the relative abundance of different sexual orientations.  One can ask, then, how the results from such studies compare with the derived results from the model of this paper.

Such a comparison should, of course, be considered to be extremely speculative in nature.  It is unlikely that the diverse range of human sexual behaviors can be described using the simplistic toy model outlined in this paper.  What's more, data on sexual orientation in humans usually divides individuals into discrete categories and relies on self-reporting of same-sex sexual behavior or sexual attraction.  All of this makes it difficult to say anything quantitative about the distribution $P(\theta)$.

\begin{figure}[htb]
\centering
\includegraphics[width=0.45 \textwidth]{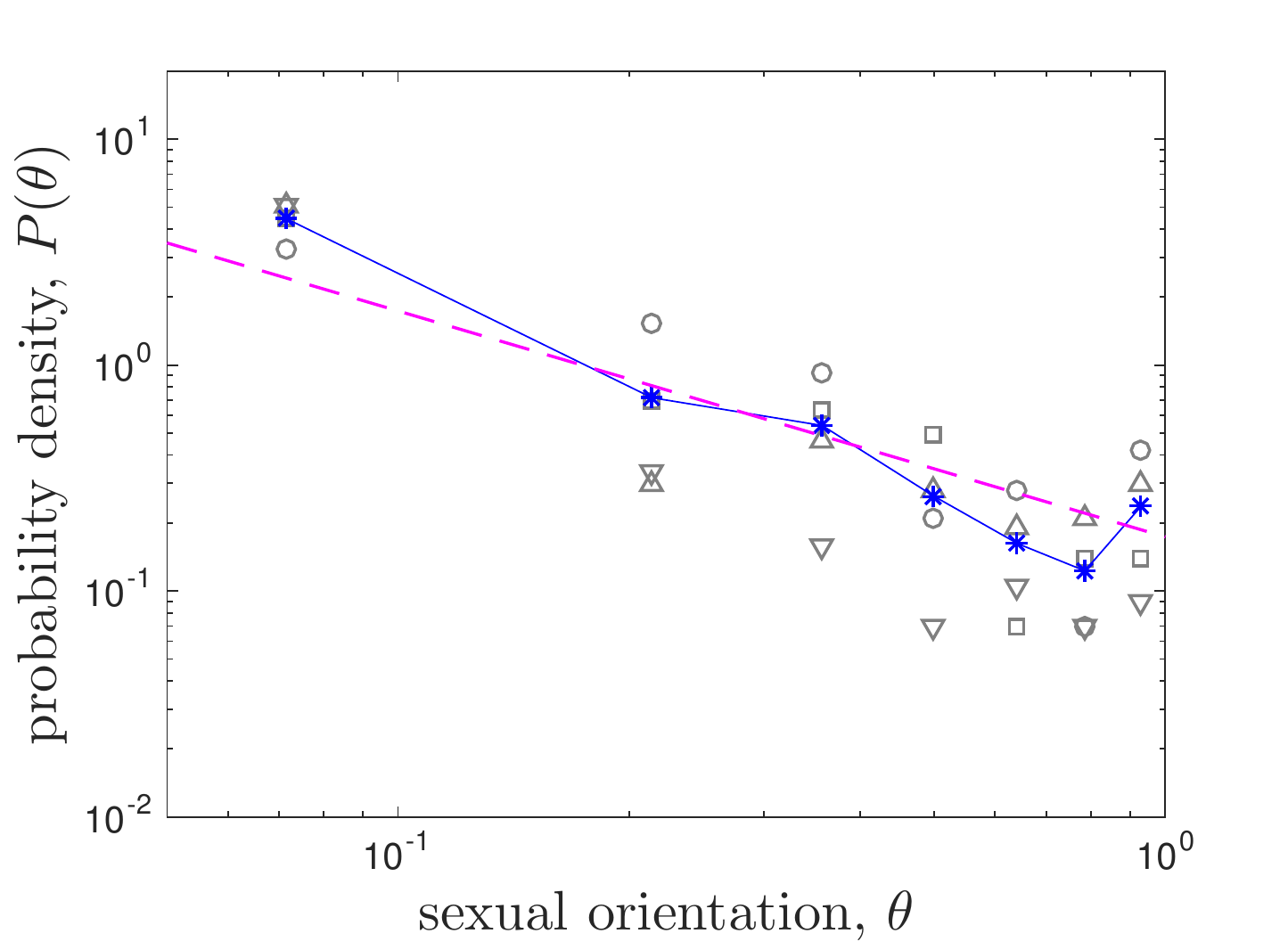}
\caption{Data for sexual orientation distribution in humans, as codified by the ``Kinsey scale", which in this plot has been uniformly spaced along the interval $0$ to $1$ and plotted in double-logarithmic scale.  Squares represent survey response data for ages 18-29 in the USA taken during the year 2015,\cite{moore_yougov_2015} circles are survey responses among ages 18-24 in the UK in 2015,\cite{dahlgreen_yougov_2015} upward-facing triangles correspond to males age 20-25 in the original Kinsey reports (published 1948),\cite{kinsey_sexual_1998} and downward-facing triangles are from females age 20-25 in the Kinsey reports (published 1953).\cite{kinsey_sexual_1998-1}  The star symbols, connected by a solid line, denote a simple average of the four data sets.  The dashed line shows a fit to Eq.\ (\ref{eq:fermiorientation}). }
\label{fig:kinsey}
\end{figure}

With these caveats, one can nonetheless make a speculative comparison between the distribution $P(\theta)$ and interview/survey data about human sexual orientation.  Such data often categorizes individuals according to their position on the Kinsey scale, which describes sexual orientation on a seven-point scale.\cite{kinsey_sexual_1998}  If this seven-point scale is (dubiously) considered to correspond to evenly-distributed intervals of the orientation $\theta$ in the range $[0, 1]$, then one can compare it directly to the theoretical distribution $P(\theta)$ from the model.  Such a comparison is presented in Fig.\ \ref{fig:kinsey}.

Figure \ref{fig:kinsey} suggests that a very approximate fit to Eq.\ (\ref{eq:fermiorientation}) is possible.  This fit gives $\Tt \approx 0.09$, which corresponds to an entropy parameter $t \approx 0.4$.  This relatively small value of $t$ is within the regime where the theoretical optimum distribution $p(t)$ is well approximated by the continuous function of Eq.\ (\ref{eq:Fermi}).  One notable failure of the model is that it is unable to capture the relatively large proportion of individuals at either extreme of the distribution, $\theta \approx 0$ and $\theta \approx 1$.  These extremes correspond to individuals who identify as either ``completely heterosexual" or ``completely homosexual", and their abundance is apparently greater than can be explained by the simple model proposed here.  It remains an interesting question whether such extremization of sexual orientation can arise from optimization of the population fitness, or whether its appearance in the data is better ascribed to other (perhaps psychological or sociological) factors.

Future and ongoing studies may allow us to adjudicate between different proposed mechanisms for the appearance of same-sex sexual behavior in the animal kingdom.  In particular, the mechanism proposed here can be refined or refuted by collecting data on the proportion $\theta$ of same-sex versus opposite-sex sexual encounters for many individuals across a large animal population, and then checking whether it obeys the characteristic $1/\theta$ distribution (as at $t \lesssim 0.5$) or whether it resembles a set of discrete delta functions (as at $t \gtrsim 1$).  

Alternatively, one could look for correlations between the rate of same-sex sexual behavior in an animal species and the diversity of expression of a particular trait.  If any such evidence is absent, it would suggest that the origins of same-sex sexual behavior cannot be described as a simple competition between increased trait diversity and increased sexual dichotomization of traits.  It could also suggest that the traits desired by a particular individual ($x_c$) are not statistically independent of the traits possessed by the individual ($x$); such non-independence would fundamentally alter the tension between the two terms in Eq.\ (\ref{eq:f}).  Either way, finding a clever way to measure and study the distribution of biological traits, $p(x)$, or the distribution of sexual orientations, $P(\theta)$, may prove to be a powerful tool for unraveling the mystery of same-sex sexual behavior.

Finally, it is worth emphasizing that many of the results presented here are specific to the quantitative form of the assumed fitness function, Eq.\ (\ref{eq:f}).  For example, one might consider that the degree of trait diversity is better characterized using the variance $\sigma^2 = \int_0^1 \left(x-\langle x \rangle \right)^2 p(x) dx$ rather than the entropy $s$. (Here, $\langle x \rangle = \int_0^1 x p(x) dx$ is the average value of the trait $x$).  Substituting $\sigma^2$ for $s$ in the fitness function $f$ would then give a different mathematical optimization problem, and thus a different trait distribution $p(x)$ for each value of the parameter $t$.  Indeed, for the specific choice $f = n(1 + t \sigma^2)$, I find that the optimal distribution $p(x)$ assumes only one of two extremes: at all $t$ larger than some critical value, $t_c \approx 25$, the optimal distribution is $p(x) = 1$, which corresponds to the $t \rightarrow \infty$ limit in Fig.\ \ref{fig:pxplot}; on the other hand, at all $t < t_c$ the optimal distribution is $2 \Theta(x - 1/2)$, where $\Theta(x)$ is the Heaviside step function, which corresponds to the $t \rightarrow 0$ limit in Fig.\ \ref{fig:pxplot}.  Thus, replacing the distribution entropy with the variance in Eq.\ (\ref{eq:f}) gives a significantly different phenomenology --- one where the population consists uniformly of either bisexual individuals ($\theta = 1/2$) or heterosexual individuals $\theta = 1$), depending on whether $t$ is above or below a critical value.

\acknowledgments 

I am grateful to several anonymous reviewers, whose comments and criticisms led to significant improvements in this manuscript.  I am also grateful to multiple friends (who probably prefer to remain anonymous), who encouraged me in this work even when I felt like its value was highly dubious.

\

\noindent \emph{Data Availability}.\; Numerical code corresponding to the analysis performed in Sec.\ \ref{sec:model} and the Appendix is available as part of this article's Supplementary Information.

\

\noindent \emph{Competing Interests}.\; I have no competing interests.

\

\noindent \emph{Author Contributions}.\; BS conceived of the study, performed the analysis, and wrote the manuscript.

\

\noindent \emph{Funding}.\; This work is unfunded.


\appendix*

\section{Numerical optimization of $p(x)$}
\label{app:numerics}

In Sec.\ \ref{sec:model} a model is introduced that relates the fitness $f$ of the population to the trait distribution $p(x)$.  Written out explicitly, this relation is
\begin{widetext}
\be 
f = \left[ \frac{1}{2} \int_0^1 \int_0^1 \frac{ \left[ (p(x_1) p(1 - x_2) \right]^2 + \left[ (p(1-x_1) p(x_2) \right]^2 }{p(x_1) p(1-x_2) + p(1-x_1) p(x_2) } dx_1 dx_2 \right] 
\times \left[ 1 - t \int_0^1 p(x) \ln p(x) dx \right].
\label{eq:flong}
\ee 
\end{widetext}
For a given value of the entropy parameter $t$, there is a specific distribution $p(x)$ that maximizes Eq.\ (\ref{eq:flong}).  This distribution can be found numerically using the following method.

First, the interval $[0, 1]$ is divided into a set of $M$ regularly-spaced points, $\{x_i\}$.  The results presented here use $M = 60$.  An initial guess is then made for the values of the distribution, $p(x_i)$, consistent with the normalization constraint
\be 
\frac{1}{M} \sum_{i = 1}^M p(x_i) = 1. \nonumber
\ee 
For the results presented in this paper, the initial guess was $p(x_i) \equiv 1$.  A Metropolis-type algorithm is then used to incrementally update the values of $p(x_i)$ in such a way that the maximum of $f$ is increasingly approached.  Specifically, the algorithm consists of repeatedly choosing random pairs of points $x_i$ and $x_j$, and then updating the values $p(x_i)$ and $p(x_j)$ such that $p(x_i) \rightarrow p(x_i) + \delta$ and $p(x_j) \rightarrow p(x_j) - \delta$.  The increment $\delta$ is chosen at random from a small interval; results presented here use $\delta \in (0, 0.01)$.  After each update, the change $\delta_f$ in the fitness is evaluated.  If $\delta_f$ is positive, then the update is kept.  If $\delta_f < 0$, on the other hand, then the update is reverted with probability $1 - \exp[\beta \delta_f]$.  Here, $\beta$ is an ``inverse temperature" parameter that determines the rate of convergence of the solution and the final numerical accuracy.

Results presented in Sec.\ \ref{sec:results} use a process of successively increasing values of $\beta$, starting at $\beta = 10^5$ and gradually increasing to $\beta = 10^{11}$.  At each value of $\beta$ a large number, $10^4 M$, of updates is attempted to ensure convergence of the solution. Care was taken to ensure that the solution converged to the same result for different random realizations of the numerical procedure.

Finally, one can notice that Eq.\ (\ref{eq:flong}) has no explicit dependence on the value of $x$, and therefore the numerical procedure does not, in general, find a set of values $\{p(x_i)\}$ that is meaningfully ordered as a function of $x_i$.  One can therefore arrange the numerical values $\{p(x_i)\}$ in order of decreasing value, and the resulting solution produces the same value of the fitness $f$.

\bibliography{orientation}

\end{document}